\documentclass[12pt]{article}

\renewcommand{\title}[1]{
\begin{center} \Large \bf #1 \end{center}
}

\renewcommand{\author}[2]{
 \begin{center} #1  \vspace{3mm} \\
  #2 \\
 \end{center}
\addvspace{\baselineskip}
}

\usepackage{amssymb}
\usepackage{amsmath}

\usepackage{amsthm}

\theoremstyle{definition}

\theoremstyle{remark}

\makeatletter
\@addtoreset{equation}{section}

\makeatother

\begin{document}

\baselineskip 5mm

%
\title{Gauge theories on 
noncommutative ${\mathbb C}P^N$
and BPS-like equations
%
}

\author{${}^1$ Akifumi Sako, ${}^2$ 
Toshiya Suzuki and~ ${}^2$
Hiroshi Umetsu }{
${}^1$ 
Department of Mathematics,
Faculty of Science Division II,\\
Tokyo University of Science,
1-3 Kagurazaka, Shinjuku-ku, Tokyo 162-8601, Japan\\ 
${}^2$
National Institute of Technology, Kushiro College\\
2-32-1 Otanoshike-nishi, Kushiro, Hokkaido 084-0916, Japan }

\vspace{1cm}

%

\abstract{ 
We give the Fock representation of a noncommutative
$\mathbb{C}P^N$ and gauge theories on it. 
The Fock representation is constructed based on star products 
given by deformation quantization with separation of variables 
and operators which act on states in the Fock space are explicitly 
described by functions of inhomogeneous coordinates on ${\mathbb C}P^N$.
Using the Fock
representation, we are able to
discuss the positivity of
Yang-Mills type actions and the minimal action principle. 
Other types of actions including the Chern-Simons term are also investigated.
BPS-like equations on noncommutative $\mathbb{C}P^1$ and
$\mathbb{C}P^2$ are derived from these actions.
There are analogies between BPS-like equations on $\mathbb{C}P^1$ and
monopole equations on ${\mathbb R}^3$, and BPS-like equations on
$\mathbb{C}P^2$ and instanton equations on ${\mathbb R}^8$.
We discuss solutions of these BPS-like equations. 
}


%

\section{Introduction}

We come across field theories on noncommutative spaces in various
situations.
For example, effective theories of D-branes with background $B$ fields are
given as gauge theories on noncommutative manifolds \cite{Seiberg:1999vs}.
Another example is the IIB matrix model \cite{Ishibashi:1996xs},
some classical solutions of which correspond to noncommutative gauge
theories.
These facts have motivated analyses of field theories on 
noncommutative spaces. 
(See, for example, review papers 
\cite{Nekrasov:2000ih,Szabo:2001kg,sako_review}.)
In particular, 
it has become increasingly important to investigate properties of 
gauge theories on various noncommutative manifolds.


In our preceding paper \cite{Sako:2012ws} we provided explicit
expressions of star products of noncommutative $\mathbb{C}P^N$ and
$\mathbb{C}H^N$, and in \cite{Maeda:2014fta} gauge theories on
noncommutative homogeneous K\"ahler manifolds are constructed by
using the deformation quantization with separation of variables \cite{Karabegov1996}.
The aim of this article is to investigate them in some details.

Here, we briefly review the deformation quantization
that is a way to realize noncommutative manifolds.
It is defined as follows.
Let $\cal F$ be a set of formal power series in $\hbar$ with
coefficients of $C^{\infty}$
functions on a Poisson manifold $M$
i.e. ${\cal F} := \left\{  f \ \Big| \ 
f = \sum_k \hbar^k f_k, ~f_k \in C^\infty (M)
\right\}$ ,
where $\hbar$ is a noncommutative parameter.
A star product is defined on ${\cal F}$ by 
\begin{eqnarray}
f * g = \sum_k \hbar^k C_k (f,g), \ \ \ \ \ f,g \in {\cal F},
\end{eqnarray}
such that the product satisfies the following conditions.
\begin{enumerate}
\item $*$ is an associative product.
\item $C_k$ is a bidifferential operator.
\item $C_0$ and $C_1$ are defined as 
\begin{eqnarray}
&& C_0 (f,g) = f g,  \\
&&C_1(f,g)-C_1(g,f) = i \{ f, g \}, \label{weakdeformation}
\end{eqnarray}
where $\{ f, g \}$ is the Poisson bracket.
\item $ f * 1 = 1* f = f$.
\end{enumerate}

In \cite{Maeda:2014fta}, noncommutative gauge theories
on homogeneous K\"ahler manifolds are constructed by
using deformation quantization with separation of variables.
The deformation quantization with separation of variables
is one of the methods to construct noncommutative K\"ahler
manifolds given by Kalabegov \cite{Karabegov1996}.
(See also \cite{Karabegov,Karabegov2011}.)
Physical quantities like gauge fields are
given as formal power series in a noncommutative parameter 
in deformation quantization,
and therefore it is difficult to discuss the positivity and boundedness of
physical quantities in general.
To justify processes of the minimal action principle and 
deriving BPS-like equations
we have to get rid of  difficulties resulting from using formal power series.
One of the way to do this is to choose an appropriate representation of
the noncommutative 
algebra on $({\cal F}, ~ *)$. 
For example, it is well-known that
the Fock representation is a good representation of noncommutative 
algebra in the Moyal ${\mathbb R}^N$.
For noncommutative ${\mathbb C}P^N$  the Fock representation 
described by using star products of Karabegov's deformation quantization
is given in \cite{Sako:2012ws,Sako:2013noa}.

In this article, we use the Fock representation 
given in \cite{Sako:2012ws,Sako:2013noa} to construct gauge theories.
By virtue of the Fock representation, we are able to prove the positivity of
Yang-Mills type actions and derive equations of motion.
We see analogies between these gauge theories on  noncommutative 
${\mathbb C}P^N$ and the gauge theories on 
noncommutative ${\mathbb R}^{N^2+2N}$.
Based on this observation, we propose BPS-like equations for gauge theories 
on noncommutative ${\mathbb C}P^1$ and ${\mathbb C}P^2$.
For the Yang-Mills-Higgs type model 
on noncommutative ${\mathbb C}P^1$, obtained BPS-like equations are
similar to the monopole equations given in the gauge-Higgs model in 
${\mathbb R}^{3}$ \cite{Bogomolny:1975de,Corrigan:1981fs}.
For the Yang-Mills type theory on noncommutative ${\mathbb C}P^2$, 
obtained BPS-like equations are
analogous to the instanton equations in ${\mathbb R}^{8}$ 
\cite{Corrigan:1982th}. 
We also discuss BPS-like equations in a gauge theory on
$\mathbb{C}P^2$ with an action which is a combination of the
Yang-Mills type action and the Chern-Simons type action.
Further, we study some solutions for these new BPS-like equations.


The organization of this paper is as follows.
In Section \ref{sect2}, we summarize preliminaries to investigate noncommutative gauge
theories.
In Section \ref{sect3}, we reformulate the Fock representation of 
${\mathbb C}P^N$ in more sophisticated  manner than in \cite{Sako:2012ws,Sako:2013noa}.
Using this Fock representation, we construct
noncommutative gauge theories and prove the 
positivity of the action functional of the gauge theories.
In Section \ref{sect4},
we derive the equations of motions, the Bianchi identities
and BPS-like equations.
Summaries are given in Section \ref{sect5}.

%

\section{Preliminaries to gauge theories on noncommutative ${\mathbb C}P^N$} \label{sect2}

In \cite{Maeda:2014fta},
gauge theories on noncommutative homogeneous K\"ahler
manifolds $M={\mathcal G}/{\mathcal H}$ are constructed.
In the theories, the K\"ahler manifolds are deformed by using 
deformation quantization with separation of variables given 
by Karabegov \cite{Karabegov1996}.
Here, we denote deformation quantization
with separation of variables when star products satisfy
\begin{equation}
 a * f = a f, ~~~~ f * b = f b,  \label{sov}
\end{equation} 
for any holomorphic function $a$ and any anti-holomorphic function $b$.


${\mathbb C}P^N$ is one of the typical 
homogeneous K\"ahler manifolds.
In the inhomogeneous coordinates $z^i ~(i=1, 2, \cdots, N)$, the K\"ahler
potential of $\mathbb{C}P^N$ is given by
\begin{align}
 \Phi = \ln \left(1+|z|^2\right), \label{phi}
\end{align}
where $|z|^2 = \sum_{k=1}^N z^k \bar{z}^k$.
The metric $(g_{i\bar{j}})$ is 
\begin{align}
 ds^2 &= 2g_{i\bar{j}}dz^id\bar{z}^j, \label{ds} \\
 g_{i\bar{j}} &= \partial_i \partial_{\bar{j}} \Phi
  = \frac{(1+|z|^2)\delta_{ij}-z^j \bar{z}^i}{(1+|z|^2)^2}, \label{metric}
\end{align}
and the inverse of the metric $(g^{\bar{i}j})$ is
\begin{align}
 g^{\bar{i}j} = (1+|z|^2)\left(\delta_{ij}+z^j\bar{z}^i\right). \label{inverse}
\end{align}
A star product in ${\mathbb C}P^N$ is given as follows
 \cite{Sako:2012ws,Sako:2013noa}:
\begin{align}
 f*g &= \sum_{n=0}^\infty c_n (\hbar) 
 g_{j_1 \bar{k}_1} \cdots g_{j_n \bar{k}_n} 
 \left(D^{j_1} \cdots D^{j_n} f\right)
 D^{\bar{k}_1} \cdots D^{\bar{k}_n} g,
 \label{Lf-cov}
\end{align}
where
\begin{align}
 c_n (\hbar) &= \frac{\Gamma(1-n+1/\hbar)}{n! \Gamma(1+1/\hbar)},~~~~ 
D^{\bar i} = g^{{\bar i} j} \partial_j ,~~~~ 
D^{i} = g^{i {\bar j} } \partial_{\bar j}.
\end{align}
This star product is constructed by using the way of the 
deformation quantization given in \cite{Karabegov1996}.
Note that the complex conjugate of $f * g$ is given as $ \overline{f * g} =
\bar{g} * \bar{f}$.
It is easily found that this star product satisfies the condition of 
deformation quantization
with separation of variables (\ref{sov}).

It should be noted 
 that first order 
differential operators in noncommutative spaces do not 
satisfy the Leibniz rule in general, 
but a linear differential operator defined by using a commutator such that
${\cal L} (f) =[P, f]_* := P*f - f*P,$ ($P, f \in C^\infty(M)[[\hbar]]$)
satisfies the Leibniz rule.
The star commutator $[P, f]_*$ includes higher
derivative terms of $f$ for a generic $P$.
We have to use first order 
differential operators as derivations 
to construct field theories 
having usual kinetic terms.
In the noncommutative K\"ahler manifolds 
deformed by deformation quantization with separation of variables, 
it is known that vector fields are inner derivations if and only if
the vector fields are the Killing vector fields 
\cite{{Muller:2004},{Maeda:2014fta}}.
Hence, field theories having usual kinetic terms 
should be constructed by using the star commutators
with the Killing potentials corresponding to the Killing vectors.
We denote the Killing vector field  
by ${\cal L}_a = \zeta_a^i\partial_i + \zeta_a^{\bar{i}} \partial_{\bar{i}}$.
Note that ``$a$" is not an index of tangent vector space of $M$
but  one of the Lie algebra of the isometry group ${\cal G}$.
The Killing vector ${\cal L}_a$'s satisfy
\begin{align}
[ {\cal L}_a , {\cal L}_b] = i
f_{abc} {\cal L}_c,
\end{align}  
where $f_{abc}$ is a structure constant of the Lie
algebra of ${\mathcal G}$.

Let us review the construction of a $U(k)$ gauge theory 
by using the Killing vector field ${\cal L}_a$ \cite{Maeda:2014fta}.
The Yang-Mills type action is given as follows.
We define a local gauge field 
${\cal A}_a \in C^\infty(M)[[\hbar]]\otimes u(k, {\mathbb C})$ as a
formal power series
\begin{align}
{\cal A}_a := \sum_{j=0}^\infty \hbar^j {\cal A}_a^{(j)},
\end{align}
and we define its gauge transformation by
\begin{align}
{\cal A}_a \rightarrow {\cal A}_a' = i U^{-1}* {\cal L}_a U
+U^{-1}* {\cal A}_a * U ,   \label{Atrans}
\end{align}
where $U$ and $U^{-1}$ are elements of 
$M_k (C^\infty(M)[[\hbar]] )$.
Note that the gauge field is defined not on the tangent space on $M$ but
on the tangent space on the isometry group ${\cal G}$ restricted into
${M}$. 
(For the case of ${\mathbb C}P^N$ the isometry group ${\cal G}$ is 
$SU(N+1)$.)
Here we put the condition 
\begin{align} \label{3_1}
 U^{\dagger} * U = \sum_{n=0}^\infty \hbar^n 
 \sum_{m=0}^n U^{(m)\dagger} * U^{(n-m)} = I, 
\end{align}
for $\displaystyle U= \sum_{j=0}^{\infty} \hbar^j U^{(j)}$.
The leading term $U^{(0)}$ is a map to $U(k)$.
We denote ${\cal U}(k)$ as a set of $U$ satisfying (\ref{3_1}).
Let us define a curvature of ${\cal A}_a$ by
\begin{align}
{\cal F}_{ab} :=
{\cal L}_a {\cal A}_b -
{\cal L}_b {\cal A}_a 
-i [ {\cal A}_a , {\cal A}_b ]_* -i f_{abc}{\cal A}_c . 
\end{align}
This
${\cal F}_{ab}$ transforms covariantly:
\begin{align}
{\cal F}_{ab} \rightarrow {\cal F}_{ab}' = U^{-1} * {\cal F}_{ab} *U. 
\label{Ftrans}
\end{align}
Therefore, we obtain the Yang-Mills type gauge invariant action;
\begin{align}
 S_g :=
 \int_{{\cal G}/{\cal H}} \mu_g ~ 
 \frac{1}{4} {\rm tr} \left(  \eta^{ac}\eta^{bd} 
 {\cal F}_{ab} * {\cal F}_{cd} \right), \label{YMaction}
\end{align}
where $\eta$ is the Killing form of ${\cal G}$, and 
$\mu_g$ is a trace density i.e.
$\int_M (f*g) \mu_g = \int_M (g*f) \mu_g$.
For ${\mathbb C}P^N$, the trace density is given by the Riemannian
volume element $\sqrt{g}$.
This gauge theory is expected to be the unique gauge theory 
that its kinetic term contains no higher derivative
and it connects to the usual Yang-Mills type theory
in the commutative limit.

%

\section{The Fock representation of noncommutative $\mathbb{C}P^N$ } \label{sect3}
The Fock representation of noncommutative $\mathbb{C}P^N$ with the star
product (\ref{Lf-cov}) was introduced in \cite{Sako:2012ws}. 
In this section, we improve its formulation and provide a more detailed description of it.

Under the star product (\ref{Lf-cov}), $z^i$ and $\partial_j \Phi =
z^{\bar{j}}/(1+|z|^2)$ satisfy the commutation relations for the
creation-annihilation operators,
\begin{equation}
 \left[ \partial_i \Phi,~ z^j \right]_* = \hbar \delta_{ij}, \qquad
  \left[z^i, ~z^j \right]_* =0, \qquad
  \left[ \partial_i \Phi, ~\partial_j \Phi \right]_* =0,
\end{equation}
and thus it would be natural to consider that functions on the
noncommutative $\mathbb{C}P^N$ are constructed from $z^i$ and
$\partial_j \Phi$. But, $z^i$ and $\partial_j \Phi =
z^{\bar{j}}/(1+|z|^2)$ are not hermitian conjugate each other, when an
inner product between functions $F, G$ is defined by
\begin{equation}
 (F, G) = \int d^Nz d^N\bar{z} \sqrt{g} ~(\bar{F}*G),
\end{equation}
where $\sqrt{g} = 1/(1+|z|^2)^{N+1}$.
Hence, we introduce another set of creation and annihilation operators, $a_i$
and $a_i^\dagger ~(i=1, 2, \dots, N)$ which are hermitian conjugate each
other, as follows;
\begin{align}
 a_i &= \frac{1}{\sqrt{\hbar}} 
 \partial_i \Phi * (1-\tilde{n} + \hbar)^{-1/2}_{*}
 = \frac{1}{\sqrt{\hbar}} (1-\tilde{n})^{-1/2}_{*} * \partial_i \Phi, \\
 a_i^\dagger &= \frac{1}{\sqrt{\hbar}} 
 (1-\tilde{n} + \hbar)_{*}^{1/2} * z^i 
 = \frac{1}{\sqrt{\hbar}} z^i * (1-\tilde{n})_{*}^{1/2},
\end{align}
where 
\begin{align}
 &\tilde{n} = z^i * \partial_i \Phi 
 = \frac{|z|^2}{1+|z|^2}, \\
 &[\tilde{n}, ~z^i]_* = \hbar z^i, \qquad 
 [\tilde{n}, ~\partial_i \Phi]_* = -\hbar \partial_i \Phi,
\end{align}
and $f_{*}^{1/2} \in {\cal F}$ is defined by 
\begin{align}
 f_{*}^{1/2} * f_{*}^{1/2} = f. \label{f1/2} 
\end{align}
The existence of $f_{*}^{1/2}$ for ${}^{\forall} f \in {\cal F}$ is
confirmed by solving (\ref{f1/2}) recursively.
Similarly, $f_{*}^{-1}$ and $f_{*}^{-1/2}$ are defined as
\begin{align}
 & f_{*}^{-1} * f = f * f_{*}^{-1} = 1 , \\
 & f_{*}^{-1/2} * f_{*}^{1/2} = f_{*}^{1/2} * f_{*}^{-1/2} = 1.
\end{align}

It can be easily seen that $a_i$ and $ a_i^\dagger$ satisfy the
commutation relations,
\begin{equation}
 \left[a_i, ~a_j^\dagger \right]_* = \delta_{ij}, \qquad
  \left[a_i, ~a_j \right]_* = 0, \qquad
  \left[a_i^\dagger, ~a_j^\dagger \right]_* = 0.
\end{equation} 
We can find that $a_i$ and $ a_i^\dagger$ are hermitian conjugates of
each other from 
the following relations which are calculated by using the definition of
star product (\ref{Lf-cov}),
\begin{align}
 & \bar{z}^i * (1-\tilde{n}) = \bar{z}^i * \frac{1}{1+|z|^2} = (1-
 \tilde{n} -\hbar) * \bar{z}^i 
 = \partial_i \Phi -\hbar \bar{z}^i, \\
 & \bar{z}^i = \partial_i \Phi * (1-\tilde{n}+\hbar)^{-1} 
 = (1-\tilde{n})^{-1} * \partial_i \Phi.
\end{align}
The number operator $n$ is defined as
\begin{align}
 n &= a^{\dag}_i * a_i
 = \frac{1}{\hbar} \tilde{n}.
\end{align}

As described in Section \ref{sect2},
commutators with the Killing potentials 
provide derivations corresponding to the Killing vector fields.
$\mathbb{C}P^N$ has the $SU(N+1)$ isometry and the Killing potentials
corresponding to the Killing vector fields, 
${\cal L}_a = -\frac{i}{\hbar} \left[P_a, ~~ \right]_* 
~(a=1, 2, \dots, N^2+2N)$, are given by
\begin{equation}
 P_a = i \left[(T_a)_{00} (z^i * \partial_i \Phi -1) -i (T_a)_{0i}
 \partial_{\bar i} \Phi -i (T_a)_{i0} \partial_i \Phi -i
 (T_a)_{ij} z^j * \partial_i \Phi \right].
 \label{KillingPot}
\end{equation}
Here $T_a$'s are bases of the fundamental representation matrices of the
Lie algebra $su(N+1)$ satisfying $[T_a, T_b] = if_{abc} T_c$ with the structure
constants $f_{abc}$, and their indices are assigned as
\begin{align*}
 T_a &= \left( \begin{array}{c|c}
			  (T_a)_{00} & (T_a)_{0j} \\ \hline
                          (T_a)_{i0} & (T_a)_{ij} \end{array} \right)
~ \in su({N+1}, {\mathbb C}) .
\end{align*}
These are represented by using the creation and annihilation operators as
\begin{align}
  P_a &= -i \left[
 (T_a)_{00}(1-\hbar n) 
 + \sqrt{\hbar} (T_a)_{0i} a_i^\dagger * (1-\hbar n)_{*}^{1/2} 
 \right. \nonumber \\
 & \hspace{15mm} \left.
 + \sqrt{\hbar} (T_a)_{i0} (1-\hbar n)_{*}^{1/2} * a_i
 + \hbar (T_a)_{ij} a_j^\dagger * a_i
 \right].
\end{align} 
Here, we used
\begin{equation}
 \partial_{\bar i} \Phi = \frac{z_i}{1 + |z|^2} 
 = z_i * (1-\tilde{n}) 
 = \sqrt{\hbar} a_i^\dagger * (1-\hbar n)_{*}^{1/2}.
\end{equation}
The Killing potentials constitute the Lie algebra $su(N+1)$,
\begin{align}
[P_a , P_b]_* &= - \hbar f_{abc} P_c.
\end{align}

In \cite{Sako:2012ws}, it is shown that $e^{-\Phi/\hbar}$ corresponds
to a vacuum state,
\begin{equation}
 |\vec{0} \rangle \langle \vec{0}| = e^{-\Phi/\hbar},
\end{equation}
which satisfies
\begin{align}
 & a_i * |\vec{0} \rangle \langle \vec{0}| 
 = \frac{1}{\sqrt{\hbar}} (1-\tilde{n})_{*}^{-1/2}*\partial_i \Phi 
 * e^{-\Phi/\hbar}= 0, \\
 & |\vec{0} \rangle \langle \vec{0}| * a_i^\dagger 
 = \frac{1}{\sqrt{\hbar}} e^{-\Phi/\hbar} * z^i * (1-\tilde{n})_{*}^{1/2}
 = 0, \\
 & \left(|\vec{0} \rangle \langle \vec{0}|\right) * 
 \left(|\vec{0} \rangle \langle \vec{0}|\right) =
 e^{-\Phi/\hbar}*e^{-\Phi/\hbar} = e^{-\Phi/\hbar}
 = |\vec{0} \rangle \langle \vec{0}|.
\end{align}
The bases of the set of linear operators acting on the Fock space are given by
\begin{align}
 &|{\vec n} \rangle \langle{\vec m}| 
 = |n_1, \cdots, n_N \rangle \langle m_1, \cdots, m_N| \nonumber \\
 &= \frac{1}{\sqrt{n_1 ! \cdots n_N !}}
 (a^{\dag}_1)_{*}^{n_1} * \cdots * (a^{\dag}_N)_{*}^{n_N} * 
 |\vec{0} \rangle \langle \vec{0}| *
 (a_1)_{*}^{m_1} * \cdots * (a_N)_{*}^{m_N} \frac{1}{\sqrt{m_1 ! \cdots m_N !}} \\
 &= \frac{1}{\sqrt{\prod^N_{i=1} n_i ! m_i !}}
 \frac{\Gamma(1/\hbar +1)}
 {\sqrt{\Gamma(1/\hbar -|n| +1)\Gamma(1/\hbar -|m| +1)}}
 \frac{\prod^N_{j=1} (z_j)^{n_j} ({\bar z}_j)^{m_j}}{(1 +
 |z|^2)^{\frac{1}{\hbar}}},
 \label{state-nm}
\end{align}
where $(a)_{*}^n = \overbrace{a * \cdots  * a}^n$ and $|m|=\sum_{i=1}^N m_i, ~|n|=\sum_{i=1}^N n_i$.
Here we used
\begin{align}
 & a_{i_1}^\dagger * \cdots * a_{i_k}^\dagger 
 = \left(\frac{1}{\sqrt{\hbar}}\right)^k
 z^{i_1} * (1-\tilde{n})^{1/2} * \cdots * z^{i_k} * (1-\tilde{n})^{1/2}
 \nonumber \\
 & 
 = z^{i_1} * \cdots * z^{i_k} 
 * (1/\hbar - {n} - (k-1))^{1/2}
 * (1/\hbar - {n} - (k-2))^{1/2} 
 * \cdots * (1/\hbar - {n})^{1/2} 
 \nonumber 
\end{align}
and 
\begin{align}
 a_{i_1}^\dagger * \cdots * a_{i_k}^\dagger * 
 |\vec{0} \rangle \langle \vec{0} |
 &= \sqrt{(1/\hbar) (1/\hbar -1) \cdots (1/\hbar-k+1)}
 z^{i_1} * \cdots * z^{i_k} * e^{-\Phi/\hbar}
 \nonumber \\
 &= \sqrt{\frac{\Gamma(1/\hbar +1)}{\Gamma(1/\hbar -k +1)}} 
 \frac{z^{i_1} \cdots z^{i_k}}{(1+|z|^2)^{1/\hbar}}.
\end{align}

We make a relation between the trace Tr on the Fock space and the
integration on $\mathbb{C}P^N$.
We normalize the trace as
\begin{equation}
 {\rm Tr} \ |{\vec n} \rangle \langle{\vec m}| = \delta_{{\vec n},{\vec
  m}}. \label{Tr_nm}
\end{equation}
Next, we calculate 
\begin{align}
I(\vec{n}, \vec{m}) = \int |{\vec n} \rangle \langle{\vec m}| ~\sqrt{g} 
\prod_{i=1}^N dz^i d\bar{z}^i.
\end{align}
First, we have
\begin{align}
 I(\vec{n}, \vec{m}) =
 & \frac{1}{\sqrt{\prod_{i=1}^N n_{i}! m_{i}!}}
 \frac{\Gamma(1/\hbar + 1)}
      {\sqrt{\Gamma(1/\hbar-|n|+1)\Gamma(1/\hbar-|m|+1)}} 
 \nonumber \\
 & \times \int \left(\prod_{i=1}^N dz^i d\bar{z}^i\right)
 \frac{\prod_{i=1}^N (z^i)^{n_i} (\bar{z}^i)^{m_i}}
      {(1+|z|^2)^{1/\hbar+N+1}}.
\end{align} 
By taking the parametrizations, $z^i = \sqrt{y_i} e^{i\theta_i}$, the
integrations by the angle variables give $(2 \pi)^N \delta_{\vec{n},\vec{m}}$.
Then $I(\vec{n}, \vec{m})$ becomes
\begin{align}
 I(\vec{n}, \vec{m}) 
 &= \delta_{\vec{n},\vec{m}}
 \frac{1}{\left( \prod_{i=1}^N n_{i}! \right)}
 \frac{(2\pi)^N \Gamma(1/\hbar + 1)}{\Gamma(1/\hbar-|n|+1)}
 \int_0^\infty \left( \prod_{i=1}^N dy^i \right)
 \frac{\prod_{i=1}^N y_i^{n_i}}
      {\left( 1+\sum_{i=1}^N y_i \right)^{1/\hbar+N+1}}.
\end{align}
The $y_i$ integrations are performed recursively from $i=1$ to $i=N$
by using the following equation,
\begin{align}
 & \int_0^\infty dy_i
 \frac{y_i^{n_i}}
 {\left(1+ \sum_{j=i}^N y_j\right)^{1/\hbar+N+1-\sum_{j=1}^{i-1} (n_j+1)}}
 \nonumber \\
 &= \frac{1}
 {\left(1+ \sum_{j=i+1}^N y_j\right)^{1/\hbar+N+1-\sum_{j=1}^i (n_j+1)}}
 \int_0^\infty dx 
 \frac{x^{n_i}}{(1+x)^{1/\hbar+N+1-\sum_{j=1}^{i-1} (n_j+1)}}
 \nonumber \\
 &= \frac{n_i! \Gamma(1/\hbar+N+1-\sum_{j=1}^i (n_j+1))}
         {\Gamma(1/\hbar+N+1-\sum_{j=1}^{i-1} (n_j+1))}
 \frac{1}
 {\left(1+ \sum_{j=i+1}^N y_j\right)^{1/\hbar+N+1-\sum_{j=1}^i (n_j+1)}},
\end{align}
where $x$ is defined as
\begin{equation}
 x = \frac{y_i}{1+\sum_{j=i+1}^N y_j}.
\end{equation}
Then we find
\begin{equation}
 I(\vec{n}, \vec{m})
 = \delta_{\vec{n}, \vec{m}} 
  c_N(\hbar ), \label{I_nm}
\end{equation}
where
\begin{equation}
c_N(\hbar )
  = 
  \frac{(2\pi)^N \Gamma(1/\hbar+1)}{\Gamma(1/\hbar+N+1)}. 
\end{equation}
The condition that all of the integrations converge is given by
\begin{equation}
 |n|, |m| < \frac{1}{\hbar} + 1. \label{cond.conv}
\end{equation}
%
%
{}From (\ref{Tr_nm}) and (\ref{I_nm}), Tr on the Fock space is related to the integration,
\begin{equation}
 {\rm Tr} \ |{\vec n} \rangle \langle{\vec m}| 
  = \frac{1}{c_N}
  \int |{\vec n} \rangle \langle{\vec m}| ~\sqrt{g} 
\prod_{i=1}^N dz^i d\bar{z}^i.
\end{equation}
 
When $1/\hbar$ is equal to a positive integer $L$, the space spanned by
the bases $|{\vec n} \rangle \langle{\vec m}|$ is consistently
restricted to those with $0 \leq |n|, |m| \leq L$. 
Then, it can be shown that the bases are complete,
\begin{align}
 \sum_{0 \leq |n| \leq L}
 |{\vec n} \rangle \langle{\vec n}|
 = \sum_{0 \leq |n| \leq L} \frac{L!}{(\prod_{i=1}^N n_i!) (L-|n|)!} 
 \frac{\prod_{i=1}^N \left(|z^i|^2\right)^{n_i}}{(1+|z|^2)^L}
 =1,
\end{align}
where $\sum_{0 \leq |n| \leq L}$ denotes the summation over all
partitions ${\vec n} =(n_1,\cdots,n_i,\cdots,n_N)$ satisfying $ 0 \leq
|n| = \sum_{i=1}^N n_i \leq L$. 

Let us consider $U(k)$ gauge theories in the Fock representation.
We here take a gauge field being anti-hermitian, and a curvature
being hermitian,
\begin{equation}
 {\cal A}_a^\dagger = - {\cal A}_a, \qquad
  {\cal F}_{ab}^\dagger = {\cal F}_{ab}.
\end{equation}
They are  expressed in the Fock representation as
\begin{align}
 {\cal A}_a &= i \sum_{\alpha,{\vec n},{\vec m}}  
 ~{\cal A}^\alpha_{a; {\vec n},{\vec m}} 
 ~t_\alpha  ~|{\vec n} \rangle \langle {\vec m}|, \\
 {\cal F}_{ab} &= \sum_{\alpha,{\vec n},{\vec m}}
 ~{\cal F}^\alpha_{ab;{\vec n},{\vec m}}
 ~t_\alpha ~|{\vec n} \rangle \langle{\vec m}|,
 \label{curvature-Fock}
\end{align}
where $t_\alpha ~(\alpha=1, 2, \dots, k^2)$ are $d \times d$ hermitian
matrices as a basis of a representation of the Lie algebra $u(k)$ and 
$~{\cal A}^\alpha_{a; {\vec n},{\vec m}} , {\cal F}^\alpha_{ab;{\vec
n},{\vec m}} \in {\mathbb C}$.
The anti-hermiticity of ${\cal A}_a$ and the hermiticity of ${\cal F}_{ab}$
lead to
\begin{equation}
 {\cal A}^{\alpha}_{a;{\vec n},{\vec m}} 
  = \overline{{\cal A}^{\alpha}_{a;{\vec m},{\vec n}}},
 \qquad
 {\cal F}^{\alpha}_{ab;{\vec n},{\vec m}} 
 = \overline{{\cal F}^{\alpha}_{ab;{\vec m},{\vec n}}}.
 \label{hermiticity}
\end{equation} 
An element of the $U(k)$ gauge transformation group ${\cal U}(k)$ 
is written as
\begin{align}
 U = \sum_{\Lambda, {\vec n},{\vec m}} U_{{\vec n},{\vec m}}^\Lambda 
 M_\Lambda |{\vec n} \rangle \langle{\vec m}|,
\end{align}
where $M_\Lambda$ are bases of $GL(d; \mathbb{C})$.
From the unitarity, $U * U^\dagger = 1$,
the following condition is imposed
\begin{align}
 \sum_{\Lambda, \Lambda',{\vec m}} ~M_{\Lambda} M_{\Lambda'}^{\dag} 
 U^\Lambda_{{\vec n},{\vec m}} 
 \overline{U^{\Lambda'}_{{\vec n'},{\vec m}}} 
 = 1_{d \times d} \delta_{{\vec n},{\vec n}'}  .
\end{align}
In short, $\sum_{\Lambda} M_{\Lambda} \otimes U^{\Lambda}_{{\vec n},{\vec
m}}$ is a unitary matrix.

In the Fock representation, 
the action functional of the gauge field is expressed as
\begin{align}
 S = \frac{c_N}{4} {\rm Tr} ~{\rm tr} ~{\cal F}_{ab} {\cal F}_{ab},
\end{align}
where Tr is the trace on the Fock space and tr is the one for 
$d\times d$ matrices.  
Here, we used the Killing form for $\mathbb{C}P^N$,
$\eta^{ab} = \delta^{ab}$.  
By using the Fock representation of the
curvature (\ref{curvature-Fock}) and the hermiticity condition of 
${\cal F}_{ab}$ (\ref{hermiticity}), we finally find 
\begin{align}
 S = \frac{c_N}{4} \sum_{\alpha, \vec{n}, \vec{m}} 
 \left| {\cal F}^\alpha_{ab;{\vec n},{\vec m}} \right|^2.
\end{align}
The action for gauge fields (\ref{YMaction}) proposed in \cite{Maeda:2014fta} 
is a formal power series in the noncommutative parameter.  In the formal
power series, one can not discuss whether the action is positive
definite or not, and thus it is not possible to use the minimum action
principle. To avoid this issue, the states was restricted within the
Fock representation, and the positive action functional was obtained by the
processes in this section.

At the end of this section, we comment about a relation between our
models and the gauge theories on the fuzzy ${\mathbb C}P^N$.
As a result of the restriction (\ref{cond.conv}), the Hilbert space of
our gauge theories becomes a finite dimensional space. So we expect our 
gauge theories are equivalent to some gauge theories on fuzzy ${\mathbb
C}P^N$ \cite{CarowWatamura:1998jn,Grosse:2004wm,Dolan:2006tx}.

%

\section{New BPS-like equations in noncommutative ${\mathbb C}P^1$ and ${\mathbb C}P^2$}
\label{sect4}
\subsection{Equations of Motion and Bianchi Identities} 
\label{EOMandBianchi}

Let us consider the action for gauge fields on ${\mathbb C}P^N$, 
(\ref{YMaction}) in the Fock representation. From a variation of the action with respective to
the gauge field,
\begin{align}
\delta S= \frac{1}{2}\int d^N z \sqrt{g} ~{\rm tr}
\left(
{\cal L}_a \delta {\cal A}_b -{\cal L}_b \delta {\cal A}_a 
-2i[\delta {\cal A}_a , {\cal A}_b ]_* -i f_{abc}\delta {\cal A}_c 
\right) * {\cal F}^{ab},
\end{align}
the equations of motion are derived as
\begin{align}
{\cal D}_b {\cal F}_{ab} 
-\frac{i}{2} f_{abc} {\cal F}_{bc} = 0. \label{EOM2}
\end{align}
${\cal D}_a = {\cal L}_a - i [ {\cal A}_a , ~ ~ ]_*$ is
the covariant derivative for the adjoint representation.
Note that the curvature is rewritten as
\begin{equation}
 {\cal F}_{ab} = i[{Q}_a, {Q}_b]_* + f_{abc} {Q}_c, \label{cuv_Q}
\end{equation}
where, using the Killing potential ${\cal P}_a = -\dfrac{i}{\hbar} P_a$, 
${Q}_a$ is defined as
\begin{align}
{Q}_a = {\cal P}_a -i {\cal A}_a .
\end{align}
From the Jacobi identities,
\begin{align}
[ {Q}_a , [ {Q}_b , {Q}_c ]_*]_* 
+ [ {Q}_b , [ {Q}_c , {Q}_a ]_*]_* 
+ [ {Q}_c , [ {Q}_a , {Q}_b ]_*]_* =0,
\end{align}
the Bianchi identities are derived as
\begin{align}
 {\cal D}_a {\cal F}_{bc} + i f_{bcd} {\cal F}_{ad} 
 + {\cal D}_b {\cal F}_{ca} + i f_{cad} {\cal F}_{bd}
 + {\cal D}_c {\cal F}_{ab} + i f_{abd} {\cal F}_{cd} 
 =0.
\label{Bianchi}
\end{align}

\subsection{New BPS-like equations on noncommutative ${\mathbb C}P^1$}
\label{4_2}
In this subsection we derive BPS-like equations on ${\mathbb C}P^1$.  
The isometry group of ${\mathbb C}P^1$ is $SU(2)$, which is a 
three-dimensional space.  Hence, it seems that there is an analogy with
gauge theories on ${\mathbb R}^3$.  
Let us introduce an adjoint scalar field $\phi$ which transforms under
a gauge transformation as  
\begin{align}
\phi \rightarrow U^{-1} * \phi * U , \ \ \ \ \ U \in {\cal U}(k),
\end{align}
and its covariant derivative is given by
\begin{align}
 {\cal D}_a \phi =  {\cal L}_a \phi - i [  {\cal A}_a , \phi ]_* .
\end{align}
Let us consider the following gauge invariant action functional with the
gauge group ${\cal U}(k)$
\begin{align}
S=\frac{1}{4} \int \mu_g {\rm tr}
( {\cal F}_{ab} * {\cal F}_{ab} 
 - 2 {\cal D}_a \phi * {\cal D}_a \phi ), \label{action-CP1}
\end{align}
where the Killing form of $su(2)$, $\eta^{ab} = \delta^{ab}$, is used.
In the Fock representation, the action is written as
\begin{align}
 & S = \frac{c_1}{4} {\rm Tr ~ tr} ({\cal F}_{ab} * {\cal F}_{ab} 
 -2 {\cal D}_a \phi * {\cal D}_a \phi). 
\end{align}
As similar to the monopole theory of the Yang-Mills-Higgs model 
on ${\mathbb R}^3$,
we can rewrite the above action as follows.
\begin{align}
 & \frac{2}{c_1} S = {\rm Tr ~ tr} 
 \left\{ \left| (i {\cal D}_a \phi \pm {\cal B}_a) \right|^2 
 \mp i {\cal L}_a ({\cal B}_a \phi + \phi {\cal B}_a) \right\},  
 \label{monopole1} 
\end{align}
where ${\cal B}_a$ is defined, using the structure constant of $su(2)$ 
$f_{abc} = \epsilon_{abc}$, as
\begin{equation}
  {\cal B}_a = \frac{1}{2} \epsilon_{abc} {\cal F}_{bc}.
\end{equation}
Here the following relations are used;
\begin{align}
 {\cal B}_a {\cal B}_a &= \frac{1}{2}{\cal F}_{ab}{\cal F}_{ab}, \\
 {\cal D}_a {\cal B}_a 
 &= \frac{1}{2} \epsilon_{abc} \left[{Q}_a, {\cal F}_{bc} \right]_*
 = \frac{1}{2} \epsilon_{abc} 
 \left[ {Q}_a,  i\left[ {Q}_b, {Q}_c \right]_*
 + \epsilon_{bcd} {Q}_d \right]_* 
 =0. \label{bianchi_cp1}
\end{align}

The second term in (\ref{monopole1}) ,
\begin{align}
\int \mu_g {\rm  tr} \ 
 \left\{ i {\cal L}_a ({\cal B}_a \phi + \phi {\cal B}_a) \right\},
\label{topological charge1}
\end{align}
is a total divergence $i \int \partial_{\mu}
\{ \mu_g~ {\rm tr}~ \xi^{\mu}_{a} (B_a
\phi + \phi B_a) \}$, in other words, topological charge.
Finally, the BPS-like equations are obtained
\begin{align}
 {\cal B}_a \pm i {\cal D}_a \phi = 0.
 \label{BPS-CP1}
\end{align}

The solutions of (\ref{BPS-CP1}) satisfy the equations of motion of (\ref{action-CP1});
\begin{align}
 & {\cal D}_a {\cal D}_a \phi = 0, \label{EOM1-CP1} \\
 & {\cal D}_b {\cal F}_{ab} - \frac{i}{2} \epsilon_{abc}{\cal F}_{bc}
 + i \left[ \phi, {\cal D}_a \phi \right]_* = 0, \label{EOM2-CP1}
\end{align} 
because
\begin{align*}
 {\cal D}_a {\cal D}_a \phi 
 &= \pm i {\cal D}_a {\cal B}_a = 0 
\end{align*}
by the Bianchi identities (\ref{bianchi_cp1}), and
\begin{align*}
 {\cal D}_b {\cal F}_{ab} - \frac{i}{2} \epsilon_{abc}{\cal F}_{bc}
 &= \epsilon_{abc} {\cal D}_b {\cal B}_c -i {\cal B}_a \\
 &= \mp \frac{i}{2} \epsilon_{abc}
 \left[{\cal D}_b, {\cal D}_c\right]_* \phi
 \mp {\cal D}_a \phi \\
 &= \mp \left[{\cal B}_a, \phi\right]_* \\
 &= -i \left[\phi, {\cal D}_a \phi \right]_*.
\end{align*}


The BPS-like equations
(\ref{BPS-CP1}) are similar to Bogomolny's 
monopole equations in the Yang-Mills-Higgs
model in ${\mathbb R}^3$, 
and the derivation processes are parallel with them \cite{Bogomolny:1975de}.
The similarities arise from the following facts: $SU(2)$ and 
$\mathbb{R}^3$ are three-dimensional spaces, and the Killing form
of $SU(2)$ used in the action (4.9) is the same as the Euclidean 
metric of $\mathbb{R}^3$.
However the equations (\ref{BPS-CP1}) are completely different from the
Bogomolny's monopole equations.
It can be seen from the following example.

For simplicity, we consider the $U(2)$ gauge theory on 
noncommutative ${\mathbb C}P^1$.
In \cite{Aoki},
Aoki, Iso and Nagao constructed a 't Hooft-Polyakov monopole
on an Fuzzy sphere.
Their solution is given by
\begin{align}
{\cal A}_a = i t_a ,
\end{align}
where $t_a$ is a some generator of the Lie algebra
corresponding to $SU(2)$ which is common to both the gauge group
and the isometry group.
The curvature of this solution vanishes i.e. ${\cal F}_{ab}=0$,
but this solution is not trivial.
(See also (\ref{F}) in the next subsection.)
Indeed it is shown that the solution has the monopole charge $-1$
in \cite{Aoki}. 
(Note that the monopole charge in  \cite{Aoki} is different from the 
topological charge (\ref{topological charge1}).)
This monopole solution is included in solutions of the above BPS-like 
equations (\ref{BPS-CP1}).
Because ${\cal B}_a = \frac{1}{2}\epsilon_{abc}{\cal F}_{bc}=0$,
if we choose $\phi =0$, 
then ${\cal A}_a = i {t}_a$ satisfies (\ref{BPS-CP1}), too.
Hence, we find this configuration is a nontrivial solution of (\ref{BPS-CP1}).

Some questions about the BPS-like equations (\ref{BPS-CP1}) arise naturally.
Are there solutions with non-zero ${\cal B}_a $?
Are there any other solutions with non-zero topological charge 
(\ref{topological charge1})?
Can we solve the BPS-like equations (\ref{BPS-CP1}) systematically?
These problems are left open.

\subsection{New BPS-like equations on noncommutative ${\mathbb C}P^2$}
In Section \ref{4_2}, we found that
there is an analogy between a gauge theory on the three-dimensional
Euclidean space and our gauge theory on noncommutative ${\mathbb C}P^1$,
because the dimension of the isometry group $SU(2)$ of noncommutative 
${\mathbb C}P^1$ is
three and the Killing form of $su(2)$ plays the role of the Euclidean metric.
It is natural to expect that this analogy extend to the case of
noncommutative ${\mathbb C}P^2$. 
Since the dimension of the isometry group $SU(3)$ of ${\mathbb C}P^2$ is
eight and the Killing form of the Lie algebra $su(2)$ is $\delta_{ab}$, 
we draw an analogy with a gauge theory on ${\mathbb R}^8$.
Similarly to the generalized instanton equations on the ${\mathbb R}^8$ 
given by Corrigan et al. \cite{Corrigan:1982th},
we try to derive new BPS equations on ${\mathbb C}P^2$ in this subsection.

We introduce $T_{abcd} ~ (a,b,c,d =1 , \dots , 8)$ 
which is completely anti-symmetric with respect to the indices 
$a,b,c,d$.
At first, we consider the case that $T_{abcd}$
is a constant, that is, $\partial_i T_{abcd} = \partial_{\bar i} T_{abcd}
= 0$.
Let us put conditions like the instanton equations,
\begin{align}
T_{abcd} {\cal F}_{cd} = 2 \lambda {\cal F}_{cd},
\label{instanton1}
\end{align} 
where $\lambda $ is a non-zero constant.
The consistency of the conditions requires that ${T_{abcd}~}'s$ need to satisfy
\begin{equation}
 T_{abcd} T_{cdef} = 2\lambda^2 
  \left(\delta_{ae}\delta_{bf} - \delta_{af}\delta_{be}\right).
\end{equation}
{}From the Bianchi identities (\ref{Bianchi}), we obtain 
\begin{align}
T_{abcd} \left({\cal D}_b {\cal F}_{cd} + i f_{bce} {\cal F}_{de}\right)
=0.
\end{align}
Using these conditions (\ref{instanton1}), these equations become
\begin{align}
{\cal D}_b {\cal F}_{ab} 
+ \frac{i}{2\lambda} T_{abcd} f_{bce} {\cal F}_{de} = 0.
\end{align}
Comparing these equations with the equations of motion (\ref{EOM2}), 
further conditions have to be imposed,
\begin{align}
T_{abcd} f_{bce} {\cal F}_{de} = -\lambda f_{abc} {\cal F}_{bc}.
\label{extra cond}
\end{align}
Hence, we find new BPS-like equations by the combinations
of (\ref{instanton1}) and (\ref{extra cond}) 
for constants $T_{abcd}$.

As similar to the case of ${\mathbb C}P^1$, 
we can construct a solution whose curvature ${\cal F}_{ab}$
vanishes.
Consider the case of $SU(3)$ gauge theory
on noncommutative ${\mathbb C}P^2$.
Since the curvature is written as (\ref{cuv_Q}),  
it is easily seen that for bases $t_a$ in a representation of $su(3)$ a gauge field
\begin{align}
{\cal A}_a = i t_a \label{exm2}
\end{align}
is a configuration of ${\cal F}_{ab}=0$ and this gauge connection is a solution of (\ref{instanton1}) and (\ref{extra cond}).

To observe that this solution gives nontrivial configurations of the gauge
fields, we consider the commutative limit of it.  For simplicity, we
take the bases of the fundamental representation as $t_a$ in
(\ref{exm2}), that is, $t_a$ is equal to $T_a$ in the Killing potential
(\ref{KillingPot}). 
In the commutative limit, the ordinary gauge fields, $A_i$ and
$A_{\bar{i}}$, are derived from ${\cal A}_a$ as follows;
\begin{align}
A_i &= -g_{i \bar{j}}\zeta^{\bar{j}}_a {\cal A}_a 
 = -i ( \partial_i P_a ) {\cal A}_a
 = ( \partial_i P_a ) T_a, \\
A_{\bar{i}} &= - g_{\bar{i}j}\zeta^{{j}}_a {\cal A}_a 
 = i ( \partial_{\bar{i}} P_a ) {\cal A}_a 
 = - ( \partial_{\bar{i}} P_a ) T_a.
\end{align}
The detailed relations between $A_{{i}}$, $A_{\bar{i}}$ and ${\cal A}_a$ are
given in \cite{Maeda:2014fta}.
After a straightforward calculation,
we obtain the curvature $F = dA + A\wedge A$ as
\begin{align}
F_{ij} &= F_{\bar{i}\bar{j}}=0 \\
F_{i \bar{j}}
&= i \left(
\begin{array}{c|c}
-\frac{1}{1+|z|^2}(g_{i\bar{j}}-\bar{z}^i z^j) 
& \partial_l g_{i \bar{j}} \\ \hline
\partial_{\bar{k}} g_{i \bar{j}} &
-\frac{z^k \bar{z}^l g_{i \bar{j}}}{1+|z|^2}
+g_{i \bar{k}}g_{l\bar{j}}(1+|z|^2)
\end{array}
\right)  . \label{F}
\end{align}
Here, we represented a matrix $K = (K_{AB}) \in M_{N+1}({\mathbb C}), ~(A, B=0, 1, \dots ,N)$ as
\begin{align}
K=\left(
 \begin{array}{c|c}
  K_{00} & K_{0l} \\ \hline
  K_{k0} & K_{kl}
 \end{array}
\right),
 \label{matrix-notation}
\end{align}
with $k, l = 1, 2, \dots, N$.
Hence, the ordinary curvature $F$ does not vanish though ${\cal F}_{ab}=0$.

The above discussion about the flat connections ${\cal F}_{ab} = 0$
is valid for noncommutative
gauge theories on $\mathbb{C}P^N$ for any $N$. 
When the isometry group $SU(N+1)$ of $\mathbb{C}P^N$ is a subgroup
of gauge group, ${\cal A}_a =i t_a$ is a flat connection 
in the meaning of ${\cal F}_{ab} = 0$, where $t_a$ is a generator of
the Lie algebra of a subgroup $SU(N+1)$ of the gauge group.  
In a case that $t_a$ is in the fundamental representation, 
the ordinary curvature $F$ has the form of (\ref{F}) in the commutative
limit and it does not vanish. 

\subsection{Another type of first order differential equations}
Next, we should like to investigate another possibility for $T_{abcd}$
in (\ref{instanton1}). In the following discussion, we consider only the
commutative $\mathbb{C}P^N$. So far, we do not succeed 
in applying the following
formulation to noncommutative $\mathbb{C}P^N$.  

Since $\mathbb{C}P^N$ is the coset space $SU(N+1)/S(U(1)\times U(N))$, 
let us consider $\mathbb{C}P^N$ embedded in $SU(N+1)$.  
Projection operators $h_{ab}$ to the tangential directions of 
$\mathbb{C}P^N$ in $SU(N+1)$ are constructed from the Killing vectors 
$\zeta_a^\mu$ of $\mathbb{C}P^N$ as
\begin{align}
 h_{ab} &= g_{\mu\nu} \zeta_a^\mu \zeta^\nu_b, \\
 h_{ac} h_{cb} &= h_{ab}.
\end{align}
From $h_{ab}$ and the completely antisymmetric covariant tensor $E_{\mu\nu\rho\sigma}$ on
$\mathbb{C}P^N$, we introduce two tensors
$J_{abcd}$ and $T_{abcd}$ as follows;
\begin{align}
 J_{ab,cd} &= \frac{1}{4}(h_{ac}h_{bd} - h_{ad}h_{bc}), \\
 T_{ab,cd} &= \frac{1}{4} 
 \zeta_a^\mu \zeta_b^\nu \zeta_c^\rho \zeta_d^\sigma
 E_{\mu\nu\rho\sigma}.
\end{align}
Our definition of $E_{\mu\nu\rho\sigma}$ is $E_{1234} = \sqrt{g}$.
These tensors satisfy the following relations,
\begin{align}
 J_{ab,ef} J_{ef,cd} &= \frac{1}{2} J_{ab,cd}, \qquad
 T_{ab,ef} T_{ef,cd} = \frac{1}{2} J_{ab,cd}, \\
 J_{ab,ef} T_{ef,cd} &= T_{ab,ef} J_{ef,cd} = \frac{1}{2} T_{ab,cd}.
\end{align}
We then define orthogonal projection operators,
\begin{align}
 P^{(\pm)}_{ab,cd} &= J_{ab,cd} \pm T_{ab,cd}, \label{pro1} \\
 P^{(\pm)}_{ab,cd} P^{(\pm)}_{cd,ef} &= P^{(\pm)}_{ab,ef}, \qquad
 P^{(\pm)}_{ab,cd} P^{(\mp)}_{cd,ef} = 0. \label{pro2}
\end{align}

Let us consider the following conditions for the curvature 
${\cal F}_{ab}$,
\begin{align}
 P^{(-)}_{ab,cd} {\cal F}_{cd} &= 0, \label{instanton}\\
 \left(I_{ab,cd} - P^{(+)}_{ab,cd} \right) 
 {\cal F}_{cd} &= 0,
\end{align}
where 
$ I_{ab,cd} = \frac{1}{2} 
(\delta_{ac} \delta_{bd} - \delta_{ad} \delta_{bc})$.  
These are equivalent to
\begin{align}
 J_{ab,cd} {\cal F}_{cd} &= \frac{1}{2} {\cal F}_{ab}, 
 \label{hhF} \\
 T_{ab,cd} {\cal F}_{cd} &= \frac{1}{2} {\cal F}_{ab}.
 \label{TF}
\end{align}
Combining these conditions and the Bianchi identities (\ref{Bianchi}), 
we find
\begin{align}
 0 &= T_{da,bc}({\cal D}_a {\cal F}_{bc} + if_{bce} {\cal F}_{ae}) 
 \nonumber \\
 &= {\cal D}_a (T_{da,bc}{\cal F}_{bc})
 -({\cal D}_a T_{da,bc}){\cal F}_{bc}
 +i T_{da,bc} f_{bce} {\cal F}_{ae} \nonumber \\
 &= \frac{1}{2} {\cal D}_a {\cal F}_{da}
 -({\cal L}_a T_{da,bc}){\cal F}_{bc}
 +i T_{da,bc} f_{bce} {\cal F}_{ae}. 
 \label{sd-1}
\end{align} 
The second and third terms in the rightest hand side of the
above equation are calculated as
\begin{align}
 -({\cal L}_a T_{da,bc}) {\cal F}_{bc}
 &= -\frac{i}{2} f_{ade} {\cal F}_{ea}
 -2i T_{da,bc} f_{bce} {\cal F}_{ae}, 
 \label{sd-2} \\
 T_{da,bc} f_{bce} {\cal F}_{ae} 
 &= 2T_{da,bc} f_{bce} T_{ae,fg} {\cal F}_{fg} \nonumber \\
 &= -\frac{1}{4} f_{bce} h_{de} h_{bf} h_{ch} {\cal F}_{fh} \label{sd-3}\\
 &= -\frac{1}{4} f_{bce} h_{de} {\cal F}_{bc}.  \notag
\end{align}
The proofs of (\ref{sd-2}) and (\ref{sd-3}) are given in Appendix \ref{appenA}.

In Appendix \ref{appendix-f}, the proof of the formulas,   
\begin{equation}
f_{abc} \zeta_a^\mu \zeta_b^\nu \zeta_c^\rho = 0, 
\label{fzzz}
\end{equation}
are given in the case of $\mathbb{C}P^N$.
Using (\ref{hhF}) and  (\ref{fzzz}), the term in the rightest
hand side of (\ref{sd-3}) vanishes
\begin{equation}
 f_{bce} h_{de} {\cal F}_{bc} 
 = f_{bce} h_{de} h_{bf} h_{ch}{\cal F}_{fh} = 0.
\end{equation}
The equations (\ref{sd-1}) finally become
\begin{equation}
 {\cal D}_b {\cal F}_{ab} -if_{abc} {\cal F}_{bc} = 0.
  \label{instanton2}
\end{equation}


%

%
%

These equations are not same as the equations of motion (\ref{EOM2}).
However, one can find a new action functional whose equations of motion
are identified with (\ref{instanton2}).

In Ref. \cite{Kawai:2010sf}, the Chern-Simons like actions on coset
spaces are provided. In our formulation, the corresponding action is written as
\begin{align}
 S_{CS} &= \int \mu_g f_{abc} 
 {\rm tr} \left(
 \frac{i}{2} {\cal A}_a * {\cal L}_b {\cal A}_c
 + \frac{1}{3} {\cal A}_a * {\cal A}_b * {\cal A}_c 
 + \frac{1}{4}f_{abd} {\cal A}_c * {\cal A}_d
 \right) \nonumber \\
 &= \int \mu_g f_{abc}
 {\rm tr} \left(
 \frac{i}{4}{\cal A}_a * {\cal F}_{bc}
 - \frac{1}{6} {\cal A}_a * {\cal A}_b * {\cal A}_c
 \right) .
\end{align}
The equations of motion of $S_{CS}$ are obtained as 
\begin{equation}
 f_{abc}{\cal F}_{bc} =0.
\end{equation}
Under the gauge transformation (\ref{Atrans}), this action transforms as
\begin{align}
 S_{CS} ~\longrightarrow~ S_{CS} 
 & + \frac{1}{2} \int \mu_G {\cal L}_a 
 \left(f_{abc} {\rm tr} ~{\cal L}_b U * U^{-1} * {\cal A}_a \right)
 \nonumber \\
 & + \frac{i}{6} \int \mu_G f_{abc} {\rm tr}~
 U^{-1} * {\cal L}_a U * U^{-1} * {\cal L}_b U * U^{-1} * {\cal L}_c U.
\end{align}
The second term on the right hand side is a total divergence, and thus
it vanishes on a compact manifold without boundary.
Furthermore, if we use the cyclic symmetry of the integration and
trace, and the fact that the Killing vectors can be written as a star
commutator, ${\cal L}_a = -\frac{i}{\hbar} [P_a, ~]_*$, it can be shown
that the third term also vanishes. Hence, $S_{CS}$ is gauge invariant.
Remark that this discussion can be applied to 
not only ${\mathbb C}P^N$ but also any homogeneous K\"ahler manifolds.

Let us consider an action which is a linear combination of the
Yang-Mills type action and the Chern-Simons type action,
\begin{equation}
 S_{YM + CS} = S_{YM} + \alpha S_{CS},
\end{equation}
with a real constant parameter $\alpha$. Then, the equations of motion of the
action are given as
\begin{equation}
 {\cal D}_b {\cal F}_{ab} 
  + \frac{i}{2}(\alpha -1) f_{abc} {\cal F}_{bc} = 0.
  \label{EOM-YMCS}
\end{equation}
Therefore, we can change the coefficient of $f_{abc} {\cal F}_{bc}$ in the
equations of motion.
If we put $\alpha = -1$, then the equations (\ref{EOM-YMCS}) 
are equal to (\ref{instanton2}).

At the end of this section, we make two comments.
First, the equations (\ref{instanton}) include usual instanton
equation $* F = \pm F$ on ${\mathbb C}P^2$ 
in the tangent direction of ${\mathbb C}P^2$.
The second comment is for the realization of this formulation in noncommutative
${\mathbb C}P^2$. To realize it, we have to construct the noncommutative
version of the projection operators (\ref{pro1}), (\ref{pro2}).
They have not been constructed until now.

\section{Conclusions}\label{sect5}

In this article, 
using the Fock representation, we investigated gauge theories on 
noncommutative ${\mathbb C}P^N$
which is constructed by means of the deformation quantization with
separation of variables. \ 
By virtue of the Fock representation, the minimal action principle
makes sense in our gauge theories.
We derived equations of motion and new BPS-like equations
for ${\mathbb C}P^1$ and ${\mathbb C}P^2$.

It is found that there are analogies between the 
gauge theories on ${\mathbb C}P^N$
and gauge theories on ${\mathbb R}^{N^2+2N}$.
Thus, the BPS-like equations on ${\mathbb C}P^1$ are similar
to the monopole equations in the Yang-Mills-Higgs model
on ${\mathbb R}^{3}$, and
the BPS-like equations on ${\mathbb C}P^2$ are similar to the 
instanton equations on ${\mathbb R}^{8}$.
%
We discussed some solutions for the BPS-like equations 
corresponding to ${\cal F}_{ab}=0$, where we note that
the vanishing of the curvature ${\cal F}_{ab}$ does not
mean that the solutions are trivial, as we saw in Section 4.
But we could not find new solutions with ${\cal F}_{ab}\neq 0$.

At the end of this article, we itemize
problems being left unsolved.\\
1. How can we solve the new BPS-like equations systematically? \\
2. Are there solutions of the BPS-like equations with ${\cal F}_{ab}\neq 0$ ? \\ 
3. How can we characterize solutions of the BPS-like equations? 
Is there any topological invariant to characterize 
and classify solutions? \\
The problem 3 is deeply related with the 
quantum corrections to the topological
term caused under the noncommutative deformation.
We set the discussion of this problem aside for another day.
\\

\noindent
{\bf Acknowledgments} \\
A.S. was supported in part by JSPS
KAKENHI Grant Number 23540117.

\appendix
\section{Proof of (\ref{sd-2}) and (\ref{sd-3})}
\label{appenA}
We let $\delta_a$ denote the Lie derivative corresponding to the Killing vector
$\zeta_a^\mu$. The Lie derivatives of the Killing vector and  
$E_{\mu\nu\rho\sigma}$ are
\begin{align}
 \delta_a \zeta_b^\mu &= \zeta_a^\nu \partial_\nu \zeta_b^\mu
 - \partial_\nu \zeta_a^\mu \zeta_b^\nu
 = if_{abc} \zeta_c^\mu, \\
 \delta_a E_{\mu\nu\rho\sigma} &= 0,
\end{align}
where the Killing equation is used.
$T_{ab,cd}$ is a scalar under the coordinate transformations, and thus
\begin{equation}
\delta_a T_{bc,de} = \zeta_a^\mu \partial_\mu T_{bc.de}
 = {\cal L}_a T_{bc,de}.
\end{equation}
Using this, (\ref{sd-2}) is obtained as follows;
      \begin{align}
       -({\cal L}_a T_{da,bc}) {\cal F}_{bc}
       &= -\frac{1}{4} \left[
       (\delta_a \zeta_d^\mu) \zeta_a^\nu \zeta_b^\rho \zeta_c^\sigma
       E_{\mu\nu\rho\sigma}
       + \zeta_d^\mu (\delta_a \zeta_a^\nu) \zeta_b^\rho \zeta_c^\sigma
       E_{\mu\nu\rho\sigma}
       + 2\zeta_d^\mu \zeta_a^\nu (\delta_a \zeta_b^\rho) \zeta_c^\sigma
       E_{\mu\nu\rho\sigma}
       \right. \nonumber \\
       & \left. \hspace{10mm}
       + \zeta_d^\mu \zeta_a^\nu \zeta_b^\rho \zeta_c^\sigma
       (\delta_a E_{\mu\nu\rho\sigma})
       \right] {\cal F}_{bc} \nonumber \\
       &= -\frac{1}{4}\left[
       if_{ade} \zeta_e^\mu \zeta_a^\nu \zeta_b^\rho \zeta_c^\sigma 
       E_{\mu\nu\rho\sigma}
       + 2i f_{abe} \zeta_d^\mu \zeta_a^\nu \zeta_e^\rho \zeta_c^\sigma
       E_{\mu\nu\rho\sigma}
       \right] {\cal F}_{bc} \nonumber \\
       &= (-i f_{ade} T_{ea,bc} -2i f_{abe} T_{da,ec}) {\cal F}_{bc}
       \nonumber \\
       &= -\frac{i}{2} f_{ade} {\cal F}_{ea}
       -2i T_{da,bc} f_{bce} {\cal F}_{ae}.
\end{align}
(\ref{sd-3}) is also shown by the following calculations.
\begin{align}
       T_{da,bc} f_{bce} {\cal F}_{ae} 
       &= 2T_{da,bc} f_{bce} T_{ae,fg} {\cal F}_{fg} \nonumber \\
       &= \frac{1}{8} f_{bce} {\cal F}_{fg}
       \zeta_d^\mu \zeta_a^\nu \zeta_b^\rho \zeta_c^\sigma
       E_{\mu\nu\rho\sigma}
       \zeta_a^{\mu'} \zeta_e^{\nu'} \zeta_f^{\rho'} \zeta_g^{\sigma'}
       E_{\mu'\nu'\rho'\sigma'} \nonumber \\
       &= -\frac{1}{8} f_{bce} {\cal F}_{fg}
       \zeta_d^\mu \zeta_b^\nu \zeta_c^\rho
       \zeta_e^{\mu'} \zeta_f^{\nu'} \zeta_g^{\rho'}
       E_{\mu\nu\rho\lambda}
       {E_{\mu'\nu'\rho'}}^\lambda \nonumber \\
       &= -\frac{1}{4} f_{bce} {\cal F}_{fg}
       \zeta_d^\mu \zeta_b^\nu \zeta_c^\rho
       \zeta_e^{\mu'} \zeta_f^{\nu'} \zeta_g^{\rho'}
       (g_{\mu\mu'}g_{\nu\nu'}g_{\rho\rho'}
       +g_{\mu\nu'}g_{\nu\rho'}g_{\rho\mu'}
       +g_{\mu\rho'}g_{\nu\mu'}g_{\rho\nu'}) \nonumber \\
       &= -\frac{1}{4} f_{bce} h_{de} {\cal F}_{bc}.
\end{align}

\section{Proof of (\ref{fzzz})}
\label{appendix-f}

Here, the notation for matrices in (\ref{matrix-notation}) is used.
We define a $(N+1) \times (N+1)$ matrix $P$ from the Killing
potentials $P_a$ of $\mathbb{C}P^N$ as
\begin{equation}
 P = P_a T_a,
\end{equation}
where $T_a$ is a generator of $su(N+1)$ in the fundamental
representation. $T_a$ is normalized so that
\begin{equation}
 (T_a)_{AB} (T_a)_{CD} = \delta_{AD}\delta_{BC}
  - \frac{1}{N+1} \delta_{AB} \delta_{CD}.
  \label{TT}
\end{equation}
Then, $P$ is written as
\begin{align}
 P = \frac{-i}{1+|z|^2}
 \left(
 \begin{array}{cc}
  1 & \bar{z}^l \\
  z^k & z^k \bar{z}^l
 \end{array}
 \right)
 + \frac{i}{N+1} 1_{N+1}.
\end{align}
The following relations hold for $P$ and derivatives of $P$;
\begin{align}
 & P \partial_i P = \frac{i}{N+1} \partial_i P, 
 \qquad \partial_i P P = -i \frac{N}{N+1} \partial_i P, \\
 & P \partial_{\bar{i}} P = -i\frac{N}{N+1} \partial_{\bar{i}} P, 
 \qquad \partial_{\bar{i}} P P = \frac{i}{N+1} \partial_{\bar i} P, \\
 & \partial_i P \partial_j P = 
 \partial_{\bar{i}} P \partial_{\bar{j}} P =0, 
 \label{dPdP} \\
 & {\rm Tr} ~\partial_i P = {\rm Tr} ~\partial_{\bar{i}} P =0, \qquad
 {\rm Tr} ~\partial_i P \partial_{\bar{j}} P = -g_{i\bar{j}}.
\end{align}
The Killing vectors of $\mathbb{C}P^N$ are represented by $\partial_i P$
and $\partial_{\bar{i}} P$ as
\begin{align}
 & \left[T_a, ~P\right] 
 = -\zeta_a^i \partial_i P -\zeta_a^{\bar{i}} \partial_{\bar{i}} P, \\
 & \zeta_a^i = -i g^{i\bar{j}} {\rm Tr} ~T_a \partial_{\bar{j}} P,
 \qquad 
 \zeta_a^{\bar{i}} = i g^{\bar{i}j}  {\rm Tr} ~T_a \partial_j P.
\end{align}

Using above relations, let us calculate 
$f_{abc} \zeta_a^{\bar{i}} \zeta_b^{\bar{j}} \zeta_c^{\bar{k}}$ ;
\begin{align}
 f_{abc} \zeta_a^{\bar{i}} \zeta_b^{\bar{j}} \zeta_c^{\bar{k}}
 &= - g^{\bar{i}l} g^{\bar{j}m} g^{\bar{k}n}
 \left({\rm Tr} ~T_a [T_b, T_c]\right) 
 \left({\rm Tr}~ T_a \partial_l P \right)
 \left({\rm Tr}~ T_b \partial_m P \right)
 \left({\rm Tr}~ T_c \partial_n P \right) \nonumber \\
 &= - g^{\bar{i}l} g^{\bar{j}m} g^{\bar{k}n}
 \left({\rm Tr} ~\left[T_c, \partial_l P\right] \partial_m P\right)
 \left({\rm Tr}~ T_c \partial_n P \right) \nonumber \\
 &= 0.
\end{align}
Here we used (\ref{TT}) in the second equality and (\ref{dPdP}) in the
third equality.
Similarly, 
$f_{abc} \zeta_a^{\bar{i}} \zeta_b^{\bar{j}} \zeta_c^k,
f_{abc} \zeta_a^{\bar{i}} \zeta_b^j \zeta_c^k$
and $f_{abc} \zeta_a^i \zeta_b^j \zeta_c^k $ 
also vanish.



\begin{thebibliography}{999}

%


\bibitem{Aoki} 
  H.~Aoki, S.~Iso and K.~Nagao,
  ``Ginsparg-Wilson relation and 't Hooft-Polyakov monopole on fuzzy 2 sphere,''
  Nucl.\ Phys.\ B {\bf 684}, 162 (2004)
  [hep-th/0312199].






\bibitem{Bogomolny:1975de}
  E.~B.~Bogomolny,
  ``Stability of Classical Solutions,''
  Sov.\ J.\ Nucl.\ Phys.\  {\bf 24} (1976) 449
   [Yad.\ Fiz.\  {\bf 24} (1976) 861].


\bibitem{CarowWatamura:1998jn} 
  U.~Carow-Watamura and S.~Watamura,
  ``Noncommutative geometry and gauge theory on fuzzy sphere,''
  Commun.\ Math.\ Phys.\  {\bf 212}, 395 (2000)
  [hep-th/9801195].

\bibitem{Corrigan:1982th} 
  E.~Corrigan, C.~Devchand, D.~B.~Fairlie and J.~Nuyts,
  ``First Order Equations for Gauge Fields in Spaces of Dimension Greater Than Four,''
  Nucl.\ Phys.\ B {\bf 214}, 452 (1983).





\bibitem{Corrigan:1981fs} 
  E.~Corrigan and P.~Goddard,
  ``An $n$ Monopole Solution With 4n-1 Degrees of Freedom,''
  Commun.\ Math.\ Phys.\  {\bf 80}, 575 (1981).


\bibitem{Dolan:2006tx}
  B.~P.~Dolan, I.~Huet, S.~Murray and D.~O'Connor,
  ``Noncommutative vector bundles over fuzzy $CP^N$ and their covariant derivatives,''
  JHEP {\bf 0707} (2007) 007
  [hep-th/0611209].




%







\bibitem{Grosse:2004wm} 
  H.~Grosse and H.~Steinacker,
  ``Finite gauge theory on fuzzy $CP^2$,''  Nucl.\ Phys.\ B {\bf 707}, 145 (2005)  [hep-th/0407089].  





\bibitem{Ishibashi:1996xs} 
  N.~Ishibashi, H.~Kawai, Y.~Kitazawa and A.~Tsuchiya,
  ``A Large N reduced model as superstring,''  Nucl.\ Phys.\ B {\bf 498}, 467 (1997)  [hep-th/9612115].  









%
\bibitem{Karabegov}
A. V. Karabegov, 
``On deformation quantization, on a Kahler manifold, associated to Berezin's
quantization,''
Funct. Anal. Appl. {\bf 30}, 142 (1996).

\bibitem{Karabegov1996}
  A.~V.~Karabegov,
  ``Deformation quantizations with separation of variables on a Kahler
  manifold,''
  Commun.\ Math.\ Phys.\  {\bf 180}, 745 (1996)
  [hep-th/9508013].




\bibitem{Karabegov2011}
  A.~V.~Karabegov,
  ``An explicit formula for a star product with separation of variables,''
  [arXiv:1106.4112 [math.QA]].


\bibitem{Kawai:2009vb} 
  H.~Kawai, S.~Shimasaki and A.~Tsuchiya,
  ``Large N reduction on group manifolds,''  Int.\ J.\ Mod.\ Phys.\ A {\bf 25}, 3389 (2010)  [arXiv:0912.1456 [hep-th]].  




\bibitem{Kawai:2010sf} 
  H.~Kawai, S.~Shimasaki and A.~Tsuchiya,
  ``Large N reduction on coset spaces,''  Phys.\ Rev.\ D {\bf 81}, 085019 (2010)  [arXiv:1002.2308 [hep-th]].  













%




\bibitem{Maeda:2014fta}
  Y.~Maeda, A.~Sako, T.~Suzuki and H.~Umetsu,
  ``Gauge theories in noncommutative homogeneous K\"ahler manifolds,''
  J.\ Math.\ Phys.\  {\bf 55}, 092301 (2014) 
  [arXiv:1403.5727 [hep-th]].

\bibitem{Muller:2004}
M. M\"uller-Bahns and N. Neumaier, 
``Invariant Star Products of Wick Type: Classification and Quantum 
Momentum Mappings,''
Lett. Math. Phys. {\bf 70}, 1 (2004).




\bibitem{Nekrasov:2000ih}
  N.~A.~Nekrasov,
  ``Trieste lectures on solitons in noncommutative gauge theories,''
  [hep-th/0011095].














\bibitem{sako_review}
A. ~Sako,
 `` Recent developments in instantons in noncommutative ${\mathbb R}^4$,''
 Adv.\ Math.\ Phys.\  {\bf 2010}(2010) , ID 270694, 28pp.

\bibitem{Sako:2012ws} 
  A.~Sako, T.~Suzuki and H.~Umetsu,
  ``Explicit Formulas for Noncommutative Deformations of $CP^N$ and $CH^N$,''  J.\ Math.\ Phys.\  {\bf 53}, 073502 (2012)  [arXiv:1204.4030 [math-ph]].  

\bibitem{Sako:2013noa} 
  A.~Sako, T.~Suzuki and H.~Umetsu,
  ``Noncommutative $CP^{N}$ and $CH^N$ and their physics,''
  J.\ Phys.\ Conf.\ Ser.\  {\bf 442}, 012052 (2013).






\bibitem{Seiberg:1999vs} 
  N.~Seiberg and E.~Witten,
  ``String theory and noncommutative geometry,''  JHEP {\bf 9909}, 032 (1999)  [hep-th/9908142].  

\bibitem{Szabo:2001kg}
  R.~J.~Szabo,
  ``Quantum field theory on noncommutative spaces,''
  Phys.\ Rept.\  {\bf 378}, 207 (2003)
  [hep-th/0109162].





\end{thebibliography}
\end{document}